\documentclass{elsartL}
\usepackage{graphicx}
\usepackage{amsmath}
\usepackage{amssymb}
\usepackage{mathrsfs}
\begin{document}

%\tikzstyle{decision} = [diamond, draw, fill=red!20, text width=7em, text badly centered, node distance=4cm, inner sep=0pt]
%\tikzstyle{block} = [rectangle, draw, fill=blue!20, text width=10em, text centered, rounded corners, minimum height=4em]
%\tikzstyle{line} = [draw, very thick, color=black!50, -latex']
%\tikzstyle{cloud} = [draw, ellipse,fill=yellow!20, node distance=2.5cm, minimum height=2em]

\begin{frontmatter}
\title{Comparative study of the two versions of the Microsoft Kinect$^{\rm TM}$ sensor in regard to the analysis of human motion}
\author[IMS]{M.J.~Malinowski},
\author[IMS]{E.~Matsinos{$^*$}}
%\author[InIT]{S.~Roth}
\address[IMS]{Institute of Mechatronic Systems, School of Engineering, Zurich University of Applied Sciences (ZHAW), Technikumstrasse 5, CH-8401 Winterthur, Switzerland}
%\address[InIT]{Institute of Applied Information Technology, School of Engineering, Zurich University of Applied Sciences (ZHAW), Steinberggasse 13, CH-8401 Winterthur, Switzerland}
%\address[IPT]{Institute of Physiotherapy, School of Health Professions, Zurich University of Applied Sciences (ZHAW), Technikumstrasse 71, CH-8401 Winterthur, Switzerland}
%\address[IMES]{Institute of Mechanical Systems, School of Engineering, Zurich University of Applied Sciences (ZHAW), Technikumstrasse 9, CH-8401 Winterthur, Switzerland}

\begin{abstract}The present paper is part of a broader programme, exploring the possibility of involving the Microsoft Kinect$^{\rm TM}$ sensor in the analysis of human motion. In this study, the output obtained from the two 
available versions of this sensor is critically examined. We demonstrate that the two outputs differ in regard to the variation of the physical quantities involved in the modelling of the human motion. As the original sensor 
has been found unsuitable for applications requiring high precision, the observed differences in the output of the two sensors call for the validation of the upgraded sensor on the basis of a marker-based system.
\end{abstract}
{$^*$}{E-mail: evangelos[DOT]matsinos[AT]zhaw[DOT]ch, evangelos[DOT]matsinos[AT]sunrise[DOT]ch}
\end{frontmatter}

\section{\label{sec:Introduction}Introduction}

Microsoft Kinect$^{\rm TM}$ (hereafter, simply `Kinect') \cite{Kinect}, a low-cost, portable motion-sensing hardware device, was developed by the Microsoft Corporation (Microsoft, USA) as an accessory to the Xbox $360$ video-game 
console (2010). The sensor is a webcamera-type, add-on peripheral device, enabling the operation of Xbox via gestures and spoken commands. In 2011, Microsoft released the software-development kit (SDK) for Kinect, thus enabling 
the development of applications in several standard programming languages. The first upgrade of the sensor (`Kinect for Windows v2'), both hardware- and software-wise, tailored to the needs of Xbox One, became available (in July 
2014) for general development and use. The present paper is part of our research programme, aiming at investigating the possibility of involving either of the Kinect sensors in the analysis of motion data of subjects walking or 
running `in place' (e.g., on commercially-available treadmills). If successful, Kinect could become an interesting alternative to marker-based systems (MBSs) in capturing data for motion analysis, one with an incontestably high 
benefit-to-cost ratio.

It is rather surprising that only one study, addressing the possibility of involving Kinect in the analysis of walking and running motion (i.e., not in a static mode or in slow motion), has appeared so far \cite{pwbn}. Using 
similar methodology to the one followed herein, the authors in that study came to the conclusion that the original sensor is unsuitable for applications requiring high precision. Of course, it remains to be seen whether any 
improvement in the overall quality of the output can be obtained with the upgraded sensor. The present study investigates the similarity of the output of the two sensors; in case of significant differences, the validation of 
the upgraded sensor, on the basis of a marker-based system (MBS), would be called for.

The detailed description of the methodology we follow herein may be found in Ref.~\cite{mmr}; we will refer to that paper whenever necessary. The present paper has been organised in five sections. In Section \ref{sec:Method}, 
we provide a short description of the hardware used in our study and give a summary of our methodology in the analysis of the human-motion data. In Section \ref{sec:Acquisition}, we give details on the experimental part of the 
study. The results, obtained from the analysis of the data of Section \ref{sec:Acquisition}, are presented in Section \ref{sec:Results}. We discuss the main conclusions of the paper in the last section.

\section{\label{sec:Method}Data acquisition and analysis}

\subsection{\label{sec:Kinect}The Kinect sensors}

In the original sensor, the skeletal data (`stick figure') of the output comprises $20$ time series of three-dimensional (3D) vectors of spatial coordinates, i.e., measurements of the ($x$,$y$,$z$) coordinates of the $20$ nodes 
which the sensor associates with the axial and appendicular parts of the human skeleton. In coronal (frontal) view of the subject (sensor view), the Kinect coordinate system is defined with the $x$ axis (medial-lateral) pointing 
to the left (i.e., to the right part of the body of the subject being viewed), the $y$ axis (vertical) upwards, and the $z$ axis (anterior-posterior) away from the sensor.

The nodes $1$ to $4$ are main-body nodes, identified as HIP\_CENTER, SPINE, SHOULDER\_CENTER, and HEAD. The nodes $5$ to $8$ relate to the left arm: SHOULDER\_LEFT, ELBOW\_LEFT, WRIST\_LEFT, and HAND\_LEFT; similarly, the nodes 
$9$ to $12$ on the right arm are: SHOULDER\_RIGHT, ELBOW\_RIGHT, WRIST\_RIGHT, and HAND\_RIGHT. The eight remaining nodes pertain to the legs, the first four to the left (HIP\_LEFT, KNEE\_LEFT, ANKLE\_LEFT, and FOOT\_LEFT), the 
remaining four to the subject's right (HIP\_RIGHT, KNEE\_RIGHT, ANKLE\_RIGHT, and FOOT\_RIGHT) leg~\footnote{The subject's left and right parts refer to what the subject perceives as the left and right parts of his/her body.}.

In the upgraded sensor, some modifications have been made in the naming (and placement) of some nodes. The original node HIP\_CENTER has been replaced by SPINE\_BASE (and appears slightly shifted downwards); the original node 
SPINE has been replaced by SPINE\_MID (and appears slightly shifted upwards); finally, the original node SHOULDER\_CENTER has been replaced by NECK (and also appears slightly shifted upwards). Five new nodes have been appended 
at the end of the list, one of which is a body node (SPINE\_SHOULDER, node $21$), whereas four nodes pertain to the subject's hands, HAND\_TIP\_LEFT ($22$), THUMB\_LEFT ($23$), HAND\_TIP\_RIGHT ($24$), and THUMB\_RIGHT ($25$).

In both versions, parallel to the captured video image, Kinect acquires an infrared image; captured with a CCD camera, this infrared image enables the extraction of information on the depth $z$. The sampling rate in the Kinect 
output (for the video and the skeletal data, for both versions of the sensor) is $30$ Hz.

\subsection{\label{sec:Stray}Stray motion}

The subject's motion on the treadmill is split into two components: the motion of the subject's `centre of mass' (CM), which should be considered as one reference point, moving synchronously with the subject's physical CM, and 
the motion of the subject's body parts relative to the CM. The coordinates of the CM are extracted from the main-body, the shoulder, and the hip nodes. Prior to further processing, the CM offsets ($x_{CM}$,$y_{CM}$,$z_{CM}$) are 
removed from the data at all times; thus, the motion is defined relative to the subject's CM. The largeness of the subject's `stray' motion on the treadmill is assessed on the basis of the root-mean-square (rms) of the 
distributions of $x_{CM}$, $y_{CM}$, and $z_{CM}$.

\subsection{\label{sec:Period}Period of the gait cycle}

Ideally, the period of the gait cycle $T$ is defined as the time lapse between successive time instants corresponding to identical postures of the human body (position and direction of motion of the human-body parts with respect 
to the CM). (Of course, the application of `identicalness' in living organisms is illusional; no two postures can ever be expected to be identical in the formal sense.) We define the period of the gait cycle as the time lapse 
between successive most distal positions $z$ of the same lower leg, which is identified herein as the ankle; one could also use the midpoints of the ankle and foot nodes, yet our past experience with the original sensor suggests 
caution with the foot nodes due to frequent artefacts, in particular in dorsal positions. The arrays of the time instants, at which the left or right lower leg is at its most distal position with respect to the instantaneous CM 
of the subject, are used in timing the motion of the left or right part of the human body.

The period of the gait cycle is related to two other quantities which are used in the analysis of motion data.
\begin{itemize}
\item The stride length $L$ is the product of the velocity $v$ and the period of the gait cycle: $L=v T$.
\item The cadence $C$ is defined as the number of steps per unit of time; a commonly-used unit is the number of steps per min.
\end{itemize}

\subsection{\label{sec:Waveforms}Waveforms}

Using the time-instant arrays from the analysis of the left and right lower-leg signals (as described in Subsection \ref{sec:Period}), each time series (involving one specific node and spatial direction) was split into one-period 
segments, which were subsequently superimposed and averaged, to yield a representative movement for the node and spatial direction over the gait cycle. Average waveforms for all nodes and spatial directions, representing the 
variation of the motion of that node (in 3D) within the gait cycle, were extracted separately for the left and right nodes of the extremities; waveforms were also extracted for the important angles introduced in Subsection 3.1 
of Ref.~\cite{mmr}. In case that left/right (L/R) information is not available (as, for example, in the case of the trunk angle), the right lower leg was used in the timing. All waveforms were subsequently $0$-centred; the removal 
of the average offsets is necessary, given that the two systems yield output which cannot be thought of as corresponding to the same anatomical locations.

The left and right waveforms yield two new waveforms, identified as the `L/R average' (LRA) and the `right-minus-left difference' (RLD); the L/R differences in the output must be investigated as we intend to use Kinect in order 
to extract the asymmetrical features in the motion.

\subsection{\label{sec:Options}Scoring options when comparing waveforms}

The similarity of corresponding waveforms is judged on the basis of five scoring options: Pearson's correlation coefficient, the Zilliacus error metric, the RMS error metric, Whang's score, and Theil's score. Assuming that a 
($0$-centred) waveform, obtained from the original sensor, is denoted by $k_i$ and the corresponding ($0$-centred) waveform from the upgraded sensor by $q_i$, the five scoring options are defined in Eqs.~(8)-(12) of Ref.~\cite{mmr} 
(where alternative ways for testing the similarity of the output of different systems are also discussed). All waveforms are obtained herein using $50$ histogram bins in the data processing.

At fixed velocity of the treadmill belt, we will investigate the variation of these scores for the nodes and spatial directions of the extremities. Tests will be performed on the LRA waveforms, as well as on those corresponding 
to the RLD waveforms. After studying the goodness of the association of the waveforms at fixed velocity, we will investigate velocity-dependent effects.

\section{\label{sec:Acquisition}Data acquisition}

The data acquisition involved one male adult (ZHAW employee), with no known motion problems, walking and running on a commercially-available treadmill (Horizon Laufband Adventure 5 Plus, Johnson Health Tech.~GmbH, Germany). The 
placement of the treadmill in the laboratory of the Institute of Mechatronic Systems (School of Engineering, ZHAW), where the experimentation took place, was such that the subject's motion be neither hindered nor influenced 
in any way by close-by objects. Prior to the data-acquisition sessions, the Kinect sensors were properly centred and aligned. The sensors were then left in the same position, untouched throughout the data acquisition.

The original sensor also provides information on the elevation (pitch) angle at which it is set. During our extensive tests, we discovered that this information is not reliable, at least for the particular device we used in our 
experimentation. To enable the accurate determination of the elevation angle of the sensor, we set forth a simple procedure. The subject stands (in the upright position, not moving) at a number of positions on the treadmill belt, 
and static measurements (e.g., $5$ s of Kinect data) at these positions are obtained and averaged. The elevation angle of the sensor may be easily obtained from the slope of the average (over a number of Kinect nodes, e.g., of 
those pertaining to the hips, knees, and ankles) ($y$,$z$) coordinates corresponding to these positions. The output data, obtained from the original sensor, were corrected (off-line) accordingly, to yield the appropriate spatial 
coordinates of the Kinect nodes in the `untilted' coordinate system. To prevent the original sensor from re-adjusting the elevation angle during the data acquisition (which is a problematic feature), we attached its body unto a 
plastic structure mounted on a tripod. Mounted on the same structure, about $10$ cm above the original sensor, was the upgraded sensor. A dedicated correction for the tilt effects in the output of the upgraded sensor (which does 
not provide information on the elevation angle) was performed as aforementioned. The data from the original sensor were corrected by (a rotation around the $x$ axis by) $6.1^\circ$, whereas those from the upgraded sensor by 
$7.2^\circ$. These corrections were confirmed by measurements taken with a laser distance meter. Interestingly, the value of the elevation angle, obtained from the SDK of the original sensor in this configuration, was only $5^\circ$.

It is worth mentioning that, as we are interested in capturing the motion of the subject's lower legs (i.e., of the ankle and foot nodes), the Kinect sensors must be placed at such a height that the number of lost lower-leg signals 
be kept reasonably small. Our past experience dictated that the Kinect sensor be placed close to the minimal height recommended by the manufacturer, namely around $2$ ft off the (treadmill-belt) floor. Placing the sensor 
higher (e.g., around the midpoint of the recommended interval, namely at $4$ ft off the treadmill-belt floor) leads to many lost lower-leg signals (the ankle and foot nodes are not tracked), as the lower leg is not visible 
by the sensor during a sizeable fraction of the gait cycle, shortly after toe-off (TO).

The Kinect sensor may lose track of the lower parts of the subject's extremities (wrists, hands, ankles, and feet) for two reasons: either due to the particularity of the motion of the extremity in relation to the position of 
the sensor (e.g., the identification of the elbows, wrists, and hands becomes problematic in some postures, where the viewing angle of the ulnar bone by Kinect is small) or due to the fact that these parts of the human body are 
obstructed (behind the subject) for a fraction of the gait cycle. Assuming that these instances remain rare (e.g., below about $3 \%$ of the available data in each time series, i.e., one frame in $30$), the missing values may be 
reliably obtained (interpolated) from the well-determined (tracked) data. Although, when normalised to the total number of the available values, the untracked signals usually appear `harmless', particular attention was paid in 
order to ensure that no node be significantly affected, as in such a case the interpolation could have yielded unreliable results.

Five velocities were used in the data acquisition: walking-motion data were acquired at $5$ km/h; running-motion data at $8$, $9$, $10$, and $11$ km/h. At each velocity setting, the subject was given $1$ min to adjust his movements 
comfortably to the velocity of the treadmill belt. The Kinect output spanned slightly more than $2$ min at each velocity. The variation of the distance between the subject and the Kinect sensors was monitored during the data 
acquisition; it ranged between about $2.5$ and $2.9$ m, i.e., well within the limits for the use of the sensors set by the manufacturer. The recording on the two measurement systems started simultaneously.

Regarding the subject's `stray' motion, the rms average values (over all velocities) of the $x_{CM}$, $y_{CM}$, and $z_{CM}$ distributions from the original sensor were: $21.4$, $16.9$, and $33.6$ mm; the corresponding 
values from the upgraded sensor were: $20.6$, $19.1$, and $32.1$ mm. Averages of the period of the gait cycle $T$, of the cadence $C$, and of the stride length $L$ are given in Table \ref{tab:Comparison}, separately for the two 
measurement systems at all velocities; the agreement between corresponding values is very satisfactory.

\section{\label{sec:Results}Comparison of the results obtained from the two Kinect sensors}

We commence with the LRA waveforms. The goodness of the association of the waveforms obtained with the two measurement systems for the eight node levels of the extremities (SHOULDER, ELBOW, WRIST, HAND, HIP, KNEE, ANKLE, and FOOT) 
and spatial directions was investigated as follows. Separately for each of the five scoring options of Section \ref{sec:Options}, velocity, and spatial direction, each node level was ranked according to the goodness of the 
association between the waveforms obtained from the two measurement systems. The node level with the worst association was given the mark of $0$, whereas the one with the best association was assigned the mark of $7$. The sum of 
the ranking scores over all velocities and scoring options yielded an $8 \times 3$ `matrix of goodness of the association' ($8$ node levels of the extremities, $3$ spatial directions); the minimal entry in this matrix may be as 
small as $0$ (which, in fact, was the overall score of the hips in the $z$ direction!), whereas the maximal value may be as large as $7 \times 5 \times 5 = 175$ (the maximal value was obtained for the foot nodes in the $z$ 
direction, a score of $173$). It was found that the nodes with the worst association (henceforth, NWA) between the two measurement systems correspond to the shoulders (in all three spatial directions) and the hips (in the $x$ 
and $z$ directions); the knees in the $y$ direction complete the first quartile of the similarity-index distribution. The average value of Pearson's correlation coefficient for the NWA was found equal to $0.464$ and the rms of 
the distribution to $0.387$.

We subsequently pursued the investigation of systematic differences in the performance of the sensors regarding: a) the upper and lower parts of the human body (i.e., upper- versus lower-extremity nodes) and b) the three spatial 
directions.
\begin{itemize}
\item To assess the similarity of the waveforms, obtained from the two measurement systems for the nodes of the upper and lower extremities, one-factor ANOVA tests were performed, separately for each of the five scoring options 
of Subsection \ref{sec:Options}, on the scores extracted at each velocity, for all upper-extremity nodes and spatial directions, and all lower-extremity nodes and spatial directions. Assuming a significance level 
$\mathrm{p}_{min} = 1 \%$ (the value which most statisticians adopt as the outset of the statistical significance), none of these tests resulted in significant effects, at any velocity; the minimal p-value from these tests was 
about $0.250$. The exclusion of the scores of the NWA does not affect the results significantly, except at $11$ km/h where the resulting p-value dropped to $3.39 \cdot 10^{-2}$ (which, nonetheless, exceeds the adopted 
$\mathrm{p}_{min}$ level). We thus conclude that the analysis cannot support any significant differences in the performance of the two sensors for the upper and lower extremities.
\item We next addressed the goodness of the association of the waveforms with regard to the three spatial directions $x$, $y$, and $z$. Although the corresponding ANOVA tests did not reveal significant effects (except at $8$ km/h, 
where the p-value was $5.19 \cdot 10^{-3}$), the p-values were generally small, i.e., close to our $\mathrm{p}_{min}$ threshold. After the exclusion of the scores of the NWA, all running-motion data yielded p-values below 
$\mathrm{p}_{min}$. The best-matching waveforms correspond to the depth $z$; the average values of Pearson's correlation coefficient in the $x$, $y$, and $z$ directions were found to be: $0.828$, $0.834$, and $0.973$, respectively.
\end{itemize}

It must be mentioned that pronounced velocity-dependent effects were observed in the overall similarity of the output of the two sensors. The association of the waveforms, obtained from the two Kinect sensors, appears to 
deteriorate with decreasing velocity; for instance, the average value (over all the nodes and spatial directions) of Pearson's correlation coefficient dropped (almost linearly) from $0.879$ (at $11$ km/h) to $0.609$ (at $5$ km/h). 
The exclusion of the scores of the NWA increases the overall similarity, leaving its velocity dependence almost intact; for instance, the average value of Pearson's correlation coefficient dropped (almost linearly) from $0.947$ 
(at $11$ km/h) to $0.788$ (at $5$ km/h). Similar conclusions were drawn after examining the results obtained with the other four scoring options. We have not found a plausible explanation of such a velocity dependence in the 
similarity of the output of the two sensors.

\begin{figure}
\begin{center}
\includegraphics [width=15.5cm] {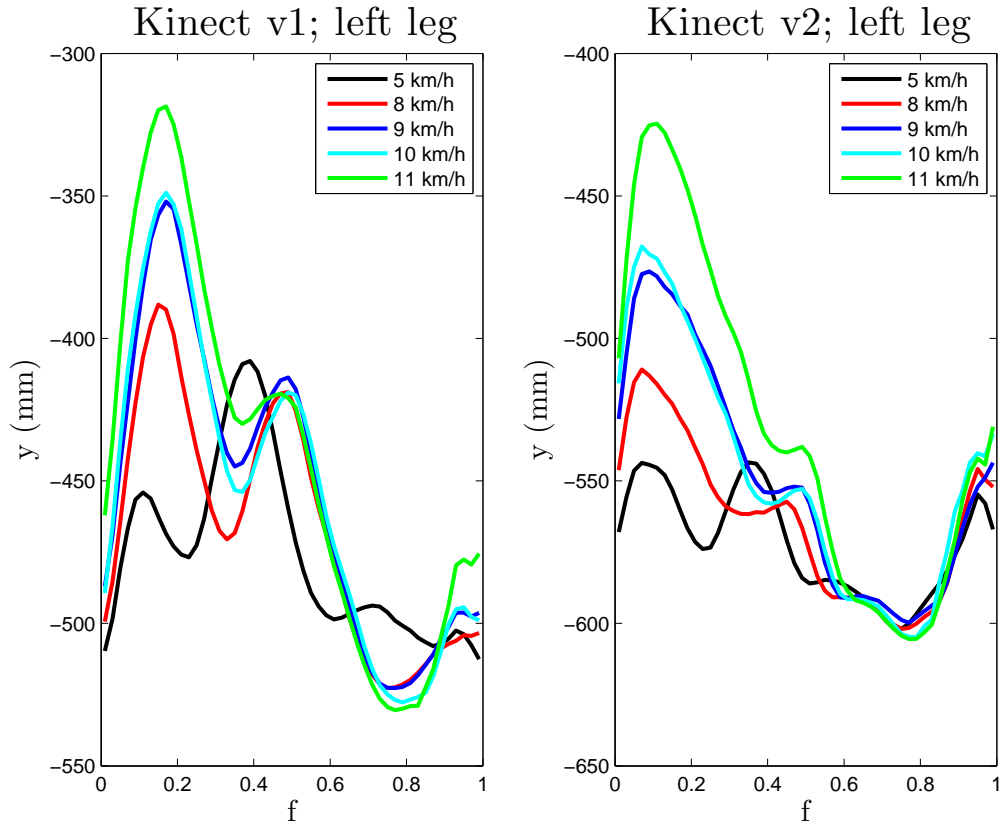}
%\vspace{-6cm}
\caption{\label{fig:LLy_L}Waveforms for the raw $y$ coordinate of the left lower leg (ankle) obtained, separately for each velocity, from the two Kinect sensors. The quantity $f$ is the fraction of the gait cycle. The difference 
in the values simply reflects the higher position of the upgraded sensor on the mount, see Section \ref{sec:Acquisition}.}
\end{center}
\end{figure}

\begin{figure}
\begin{center}
\includegraphics [width=15.5cm] {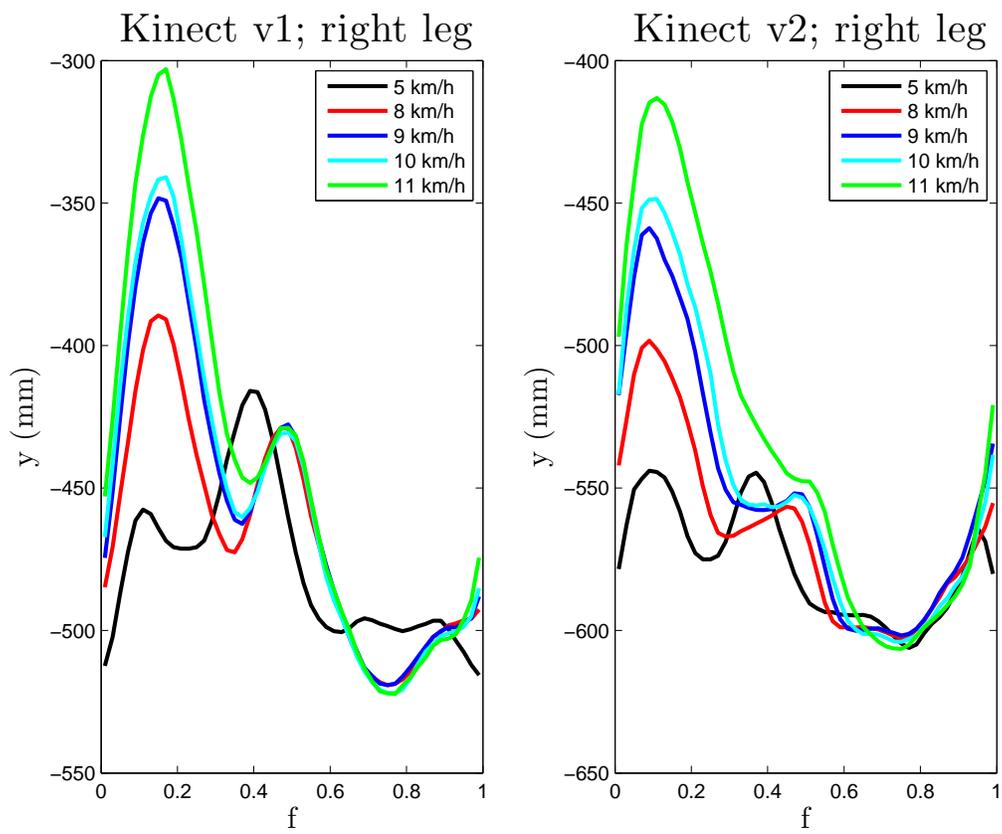}
%\vspace{-6cm}
\caption{\label{fig:LLy_R}Same as Fig.~\ref{fig:LLy_L} for the right lower leg (ankle).}
\end{center}
\end{figure}

We also compared the $y$ waveforms of the lower legs. This comparison is interesting for two reasons. First, the lower-leg signals are used in timing the motion; second, we intend to use these signals in order to obtain the times 
(expressed as fractions of the gait cycle) of the initial contact (IC) and the TO \cite{o,n}; the difference of these two values is the stance fraction. Shown in Figs.~\ref{fig:LLy_L} and \ref{fig:LLy_R} are the waveforms obtained 
from the outputs of the two measurement systems, separately at each velocity. The waveforms represent the variation of the raw signals, i.e., the $y$ offsets of the subject's CM have not been removed. The salient feature in the 
waveforms obtained with the original sensor is a pronounced peak appearing around the IC; this peak is reduced in the waveforms obtained with the upgraded sensor.

We also investigated the goodness of the association of the RLD waveforms. To this end, two-sided t-tests were performed on the score distributions between corresponding LRA and RLD waveforms, a total of $75$ tests: five scoring 
options, three tests per scoring option (paired, homoscedastic, and unequal-variance), and five velocities. The p-values, obtained from the majority of these tests for the running-motion data, were found to be small, below 
$\mathrm{p}_{min}$; the median p-value was equal to $2.36 \cdot 10^{-3}$, whereas the minimal one was $2.53 \cdot 10^{-5}$. (After the exclusion of the scores of the NWA, the median p-value dropped to $1.10 \cdot 10^{-3}$, whereas 
the minimal one to $5.69 \cdot 10^{-6}$.) The analysis showed that Pearson's correlation coefficients of the tests on the RLD waveforms were systematically below those of the LRA waveforms, whereas all other scores were larger 
for the RLD waveforms, thus indicating a poorer association of the waveforms obtained from the two measurement systems for the RLDs compared to the LRAs. As a result, the similarity between the RLD waveforms, obtained with the 
two Kinect sensors, is weaker than between the LRA waveforms.

\begin{figure}
\begin{center}
\includegraphics [width=15.5cm] {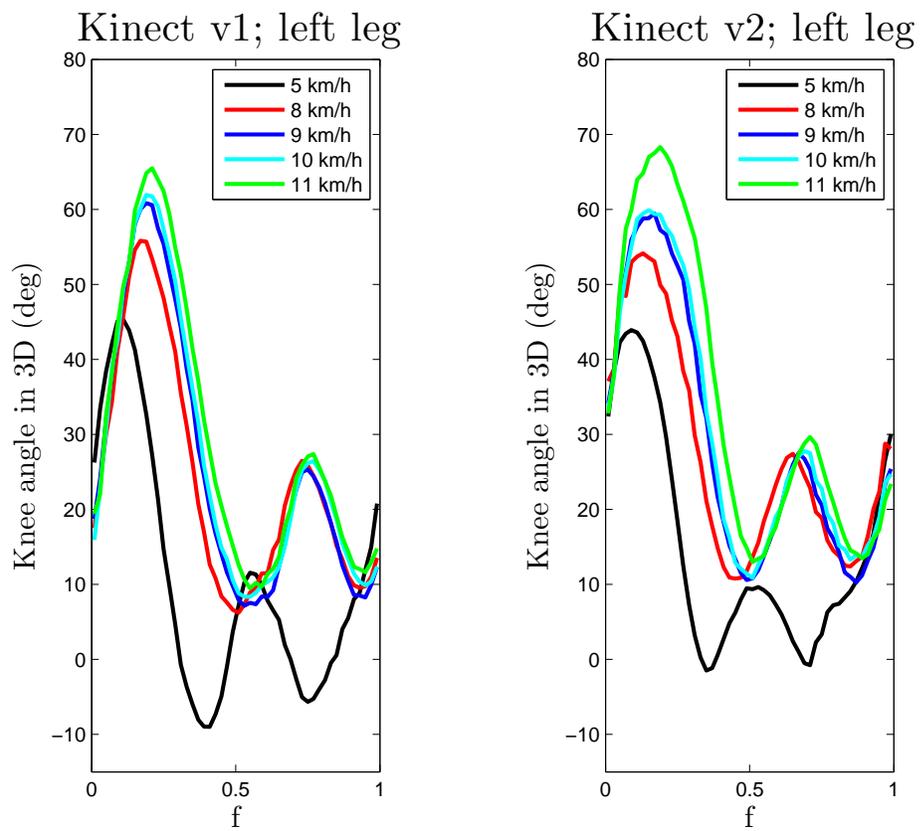}
%\vspace{-6cm}
\caption{\label{fig:ka3D_L}The waveforms for the left-knee angle in 3D obtained from the two Kinect sensors. The quantity $f$ is the fraction of the gait cycle.}
\end{center}
\end{figure}

\begin{figure}
\begin{center}
\includegraphics [width=15.5cm] {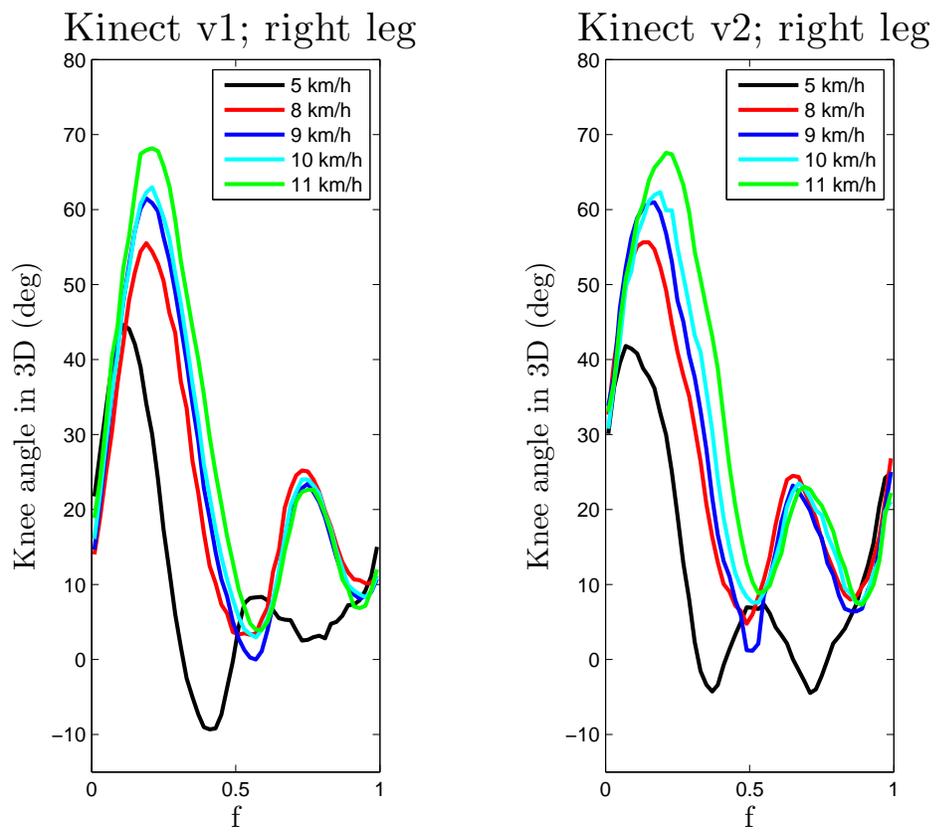}
%\vspace{-6cm}
\caption{\label{fig:ka3D_R}Same as Fig.~\ref{fig:ka3D_L} for the right knee.}
\end{center}
\end{figure}

We also examined the similarity of the waveforms for the important angles, introduced in Subsection 3.1 of \cite{mmr}. The only waveforms which match well are those for the hip and knee angles in the sagittal plane, and (to a 
lesser degree) those for the hip angles in the coronal plane. In the sagittal plane, the average values of Pearson's correlation coefficients for the LRAs were equal to $0.972$ and $0.920$ for the hip and knee angles, respectively. 
In the coronal plane, the average value of Pearson's correlation coefficient for the LRAs was equal to $0.873$ for the hip angles. The corresponding RLD waveforms do not much well; in the sagittal plane, the average values of 
Pearson's correlation coefficients were equal to $0.253$ and $0.457$; the agreement for the hip angles in the coronal plane was found to be even poorer (Pearson's correlation coefficient came out around $0.293$). The average value 
of Pearson's correlation coefficients for the LRAs in the case of the knee angle in 3D was $0.918$; for the corresponding RLDs, $0.314$; the left and right knee angles in 3D are shown in Figs.~\ref{fig:ka3D_L} and \ref{fig:ka3D_R}.

\begin{figure}
\begin{center}
\includegraphics [width=15.5cm] {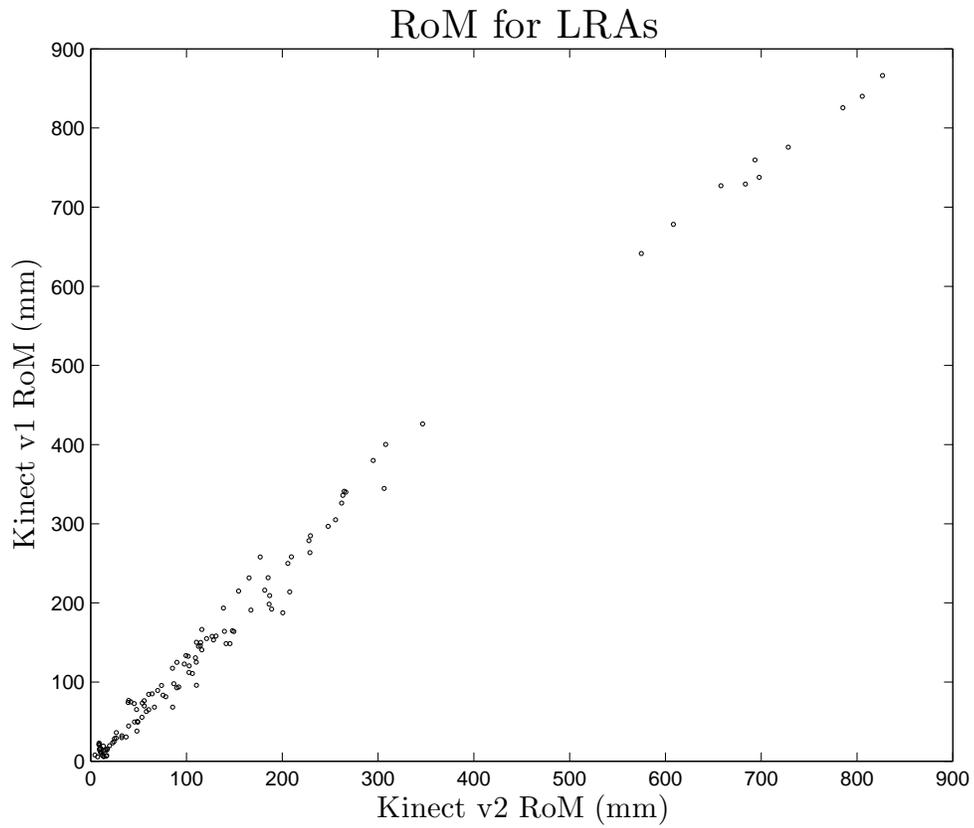}
%\vspace{-6cm}
\caption{\label{fig:RoMLRAs}The ranges of motion (RoMs) of the left/right average (LRA) waveforms, obtained with the original Kinect sensor, plotted versus those obtained with the upgraded Kinect sensor, at all velocity settings 
used in the data acquisition (see Section \ref{sec:Acquisition}).}
\end{center}
\end{figure}

\begin{figure}
\begin{center}
\includegraphics [width=15.5cm] {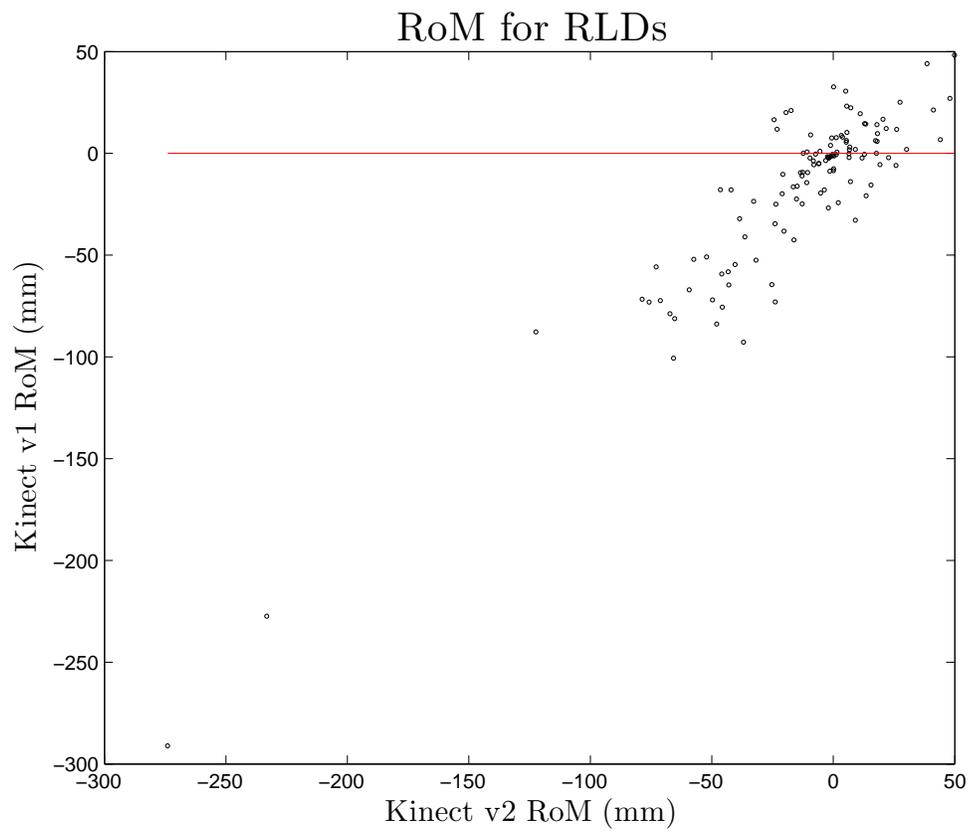}
%\vspace{-6cm}
\caption{\label{fig:RoMRLDs}Same as Fig.~\ref{fig:RoMLRAs} for the `right-minus-left' difference (RLD) waveforms.}
\end{center}
\end{figure}

We finally address the comparison of the RoMs obtained from the waveforms of two measurement systems waveforms. It might be argued that one could simply use in a study the RoMs, rather than the waveforms, as representative of 
the motion of each node. Of course, given that each waveform is essentially replaced by one number, the information content in the RoMs is drastically reduced compared to that contained in the waveforms. Shown in 
Fig.~\ref{fig:RoMLRAs} is a scatter plot of the RoMs of the LRA waveforms. The ideal relation between these two quantities is a straight line with slope equal to $1$; Pearson's correlation coefficient between the two RoM arrays 
came out equal to $0.995$. Shown in Fig.~\ref{fig:RoMRLDs} is a scatter plot of the RoMs of the RLD waveforms. A straight line with slope equal to $1$ is the expected relation here too; however, the scattering of the values is 
larger and Pearson's correlation coefficient dropped to $0.917$. Again, we draw the conclusion that the similarity of the RLD waveforms, obtained with the two Kinect sensors, is smaller to that of the LRA waveforms. An unweighted 
quadratic fit, constrained to pass through the origin of the plot, to the RoMs of the LRA waveforms resulted in curvature and slope values equal to $(-25.7 \pm 2.6) \cdot 10^{-5}$~mm$^{-1}$ and $1.259 \pm 0.016$, respectively, 
indicating that (in comparison to the upgraded sensor) the original sensor overestimates (on average) the RoMs (in the domain of the values of Fig.~\ref{fig:RoMLRAs}) by about $12 \%$. These results do not change significantly 
after the exclusion of the scores of the NWA.

\section{\label{sec:Conclusions}Discussion and conclusions}

The present study is part of a broader programme, aiming at investigating the involvement of the Microsoft Kinect$^{\rm TM}$ (hereafter, `Kinect') sensor in the analysis of human motion. In this work, we set on assessing the 
similarity in the performance of the two (original and upgraded) versions of the sensor for data acquired from one subject walking and running on a commercially-available treadmill. An estimate of the degree of the association 
of the output of these systems is obtained after following the methodology of Ref.~\cite{mmr} and comparing the waveforms extracted for the nodes of the extremities from the two measurement systems, as well as of those for the 
important angles (defined in Subsection 3.1 therein). Our comparative study of the two Kinect sensors identifies the similarities and the differences in their performance, but (of course) cannot enable conclusions regarding 
which of the two sensors performs better.

In this work, we came up with a number of significant differences in the performance of the two sensors.
\begin{itemize}
\item The worst association is obtained for the waveforms of the shoulders (in all three spatial directions), of the hips (in the $x$ and $z$ directions), and of the knees (in the $y$ direction).
\item The association of the waveforms in the two lateral directions ($x$ and $y$) is poorer than that in the $z$ direction (depth).
\item The association of the waveforms deteriorates with decreasing velocity of the walker/runner.
\item The $y$ waveforms of the lower legs come out different. A characteristic peak, which is present in the data of the original sensor shortly before the initial contact, appears reduced in the data of the upgraded sensor.
\item Exempting the hip and knee angles in the sagittal plane, and the lateral hip angles in the coronal plane, the association of the waveforms of the important angles is poor.
\item The association of the ranges of motion (RoMs) is poor. It appears that the original sensor overestimates the RoMs (in the domain of the values of the present paper) by about $12 \%$.
\item The association between the waveforms, entering the investigation of the asymmetry of the motion, is poorer than the results obtained for the average waveforms for the left and right parts of the human body.
\end{itemize}

As the original sensor has been proven unsuitable for applications requiring high precision \cite{pwbn}, it must be investigated whether the differences detailed herein constitute an improvement in the performance of the upgraded 
sensor. As a result, a dedicated validation study for the upgraded sensor, on the basis of a marker-based system, is called for.

\begin{ack}
We acknowledge several helpful discussions with S.~Roth.\\
\end{ack}

{\bf Conflict of interest statement}

The authors certify that, regarding the material of the present paper, they have no affiliations with or involvement in any organisation or entity with financial or non-financial interest.

\newpage
\begin{table}[h!]
{\bf \caption{\label{tab:Comparison}}}The average values of the period of the gait cycle $T$, of the cadence $C$, and of the stride length $L$ at the five velocity settings used in the data acquisition (see Section 
\ref{sec:Acquisition}), separately for the two Kinect sensors.
\vspace{0.6cm}
\begin{center}
\begin{tabular}{|c|c|c|c|c|c|}
\hline
 & $5$ km/h & $8$ km/h & $9$ km/h & $10$ km/h & $11$ km/h \\
\hline
\multicolumn{6}{|c|}{Original Kinect sensor} \\
\hline
$T$ (s) & $1.0910(13)$ & $0.7391(17)$ & $0.7156(23)$ & $0.69544(78)$ & $0.6834(18)$ \\
$C$ (steps/min) & $109.99(13)$ & $162.36(37)$ & $167.69(54)$ & $172.55(19)$ & $175.60(46)$ \\
$L$ (m) & $1.5152(17)$ & $1.6425(37)$ & $1.7890(57)$ & $1.9318(22)$ & $2.0880(55)$ \\
\hline
\multicolumn{6}{|c|}{Upgraded Kinect sensor} \\
\hline
$T$ (s) & $1.0896(23)$ & $0.7384(19)$ & $0.7160(19)$ & $0.6950(12)$ & $0.6826(17)$ \\
$C$ (steps/min) & $110.13(23)$ & $162.52(42)$ & $167.61(45)$ & $172.66(29)$ & $175.80(44)$ \\
$L$ (m) & $1.5134(32)$ & $1.6409(42)$ & $1.7899(48)$ & $1.9305(33)$ & $2.0858(52)$ \\
\hline
\end{tabular}
\end{center}
\end{table}

\clearpage
% ============= FIGURE 1

\clearpage
% ============= FIGURE 2

\clearpage
% ============= FIGURE 3

\clearpage
% ============= FIGURE 4

\clearpage
% ============= FIGURE 5

\clearpage
% ============= FIGURE 6

\end{document}